# Direct observation of avalanche scintillations in a THGEM-based two-phase Ar avalanche detector using Geiger-mode APD

A. Bondar[a], A. Buzulutskov[a][*], A. Grebenuk[a], A. Sokolov[a], D. Akimov[b], I. Alexandrov[b] and A. Breskin[c]

[a]*Budker Institute of Nuclear Physics, Lavrentiev avenue 11, 630090 Novosibirsk, Russia*
[b]*Institute for Theoretical and Experimental Physics, Bolshaya Cheremushkinskaya 25, 117218 Moscow, Russia*
[c]*Weizmann Institute of Science, 76100 Rehovot, Israel*
*E-mail:* `A.F.Buzulutskov@inp.nsk.su`

ABSTRACT: A novel concept of optical signal recording in two-phase avalanche detectors, with Geiger-mode Avalanche Photodiodes (G-APD) is described. Avalanche-scintillation photons were measured in a thick Gas Electron Multiplier (THGEM) in view of potential applications in rare-event experiments. The effective detection of avalanche scintillations in THGEM holes has been demonstrated in two-phase Ar with a bare G-APD without wavelength shifter, i.e. insensitive to VUV emission of Ar. At gas-avalanche gain of 400 and under ±70º viewing-angle, the G-APD yielded 640 photoelectrons (pe) per 60 keV X-ray converted in liquid Ar; this corresponds to 0.7 pe per initial (prior to multiplication) electron. The avalanche-scintillation light yield measured by the G-APD was about 0.7 pe per avalanche electron, extrapolated to $4\pi$ acceptance. The avalanche scintillations observed occurred presumably in the near infrared (NIR) where G-APDs may have high sensitivity. The measured scintillation yield is similar to that observed by others in the VUV. Other related topics discussed in this work are the G-APD's single-pixel and quenching resistor characteristics at cryogenic temperatures.

---

[*] Corresponding author.

# Contents

## 1. Introduction

In the concept of a cryogenic two-phase avalanche detector [1],[2],[3] the dense noble-liquid medium is combined with a gas avalanche multiplier: electrons extracted from liquid through the liquid-gas interface are effectively multiplied with Gas Electron Multipliers (GEMs) [4] or thick GEMs (THGEMs) [5]. This followed the remarkable property of hole-type multipliers to operate in noble gases with high gains [6],[7], due to considerably reduced avalanche-induced secondary effects. The development of such two-phase avalanche detectors has been motivated by applications in rare-event experiments and in medical imaging: in coherent neutrino-nucleus scattering [8], dark matter search [9], solar neutrino [10] and large scale neutrino [11] detectors and Positron Emission Tomography [3]. In particular for the first two applications, a low-noise self-triggering detector is desirable, with ultimate sensitivity, preferably operating in single-electron counting mode. Most promising results were obtained with two-phase Ar avalanche detectors providing gains reaching $10^4$ with a triple-GEM [12],[13],[14] and gains reaching $3\times10^3$ and several hundreds with a double-THGEM [15] and a single-THGEM [15],[16] respectively.

    In this work we investigated a novel technique of signal recording in two-phase avalanche detectors; it consists of optically recording avalanche-induced scintillation light emitted from THGEM holes, using Geiger-mode Avalanche Photodiodes (G-APDs, [17]). In detectors requiring ultimate sensitivities, the optical readout might be preferable as compared to charge readout in terms of overall gain and noise. Indeed, the typical gains in hole-multipliers listed above, might not be sufficient for single-electron counting, recording avalanche-charge in self-triggering mode. On the other hand, even at low gas-avalanche gains, the high G-APD gain, reaching $10^6$ [17], would substantially increase the overall gain, thus providing effective single-

electron counting. In addition, multi-channel optical readout with overlapping fields-of-view and coincidence between channels, would effectively suppress single-channel noise.

Moreover, the G-APD performance at cryogenic temperatures is superior to that at room temperature [18],[19],[20],[21],[22],[23],[24],[25]: the noise-rate is considerably reduced [18],[20],[24],[25], while the amplitude resolution [21] and the maximum gain [24] can be substantially increased.

Earlier studies of optical readout from THGEM in two-phase Ar were performed in [16]: the G-APD employed, coated with a wavelength shifter (WLS), was sensitive to VUV scintillations. The G-APD yielded 200 photoelectrons signals at THGEM gains of 100, for 6 keV X-rays converted in liquid Ar.

The current hypothesis regarding scintillation in Ar was that it occurs mostly in VUV, peaked at 128 nm (see for example [26]), necessitating the use of WLS coatings on visible-range photodetectors. On the other hand, it was reported [27] that Ar has intense avalanche-induced scintillation in the near-infrared (NIR), between 750 and 850 nm: of about one NIR photon per avalanche electron over $4\pi$. In this spectral region the G-APD photon detection efficiency (PDE) can be as high as 15-25% [28],[29], which would provide direct and effective detection of Ar scintillation without WLS.

Accordingly in the current work, in contrast to [16], we investigated bare G-APDs for the optical readout of the THGEM. Preliminary results were presented in [24], focused mainly on G-APD characteristics at cryogenic temperatures and avalanche-scintillation recording in two-phase Ar with a G-APD and low-gain (~60) THGEM. In the current work we present more elaborated results, including those on direct observation of avalanche scintillations in two-phase Ar using uncoated G-APD at higher THGEM gains, of ~400. The THGEM/G-APD multiplier yield, avalanche light yield, as well as electron emission and avalanche mechanisms in two-phase Ar, are described in detail. Other related topics discussed in this work are the G-APD's single-pixel and quenching resistor characteristics at cryogenic temperatures.

## 2. Experimental setup

Experimental setups and procedures for investigating our first-generation of two-phase avalanche detectors were described elsewhere [12],[13],[14],[15]. A new experimental setup was recently assembled at BINP, with larger volume (9 l) cryogenic chamber, larger X-ray windows and better temperature stability. In the two-phase mode, the chamber comprised a cathode mesh at the bottom, immersed in a ~1 cm thick liquid Ar layer and double-THGEM assembly placed within the saturated vapor above the liquid: see Fig. 1. The detector was operated in two-phase Ar in equilibrium state, at a saturated vapour pressure of 1.0 atm corresponding to a temperature of 87 K. In the cathode gap the liquid and gas layer thicknesses were 7 and 3.5 mm respectively.

Ar was taken from the bottle with a specified purity of 99.998% ($N_2$ content <0.001%); during cooling procedures it was additionally purified from oxygen and water by Oxisorb filter, providing electron life-time in the liquid >20 μs [14],[30].

The THGEMs (by *Print Electronics, Israel*) were made of G10 and had the following geometrical parameters [15]: 0.4 mm dielectric thickness, 0.9 mm hole pitch, 0.5 mm hole diameter, 0.1 mm hole rim and 25×25 mm$^2$ active area. The cathode and THGEM electrodes were biased through a resistive high-voltage divider, shown in Fig. 1, placed outside the cryostat. The grid behind the second THGEM was at floating potential. In the two-phase mode in all the measurements, the electric field within liquid Ar was 1.8 kV/cm.

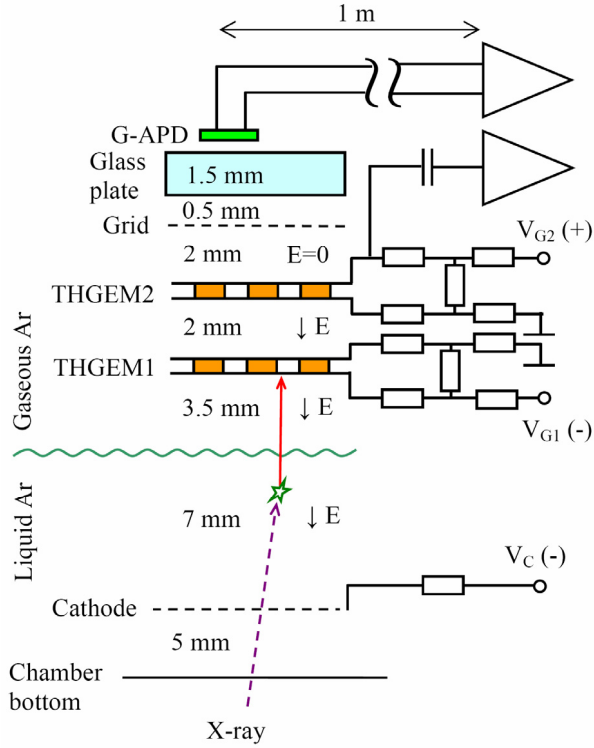

Fig. 1. Schematic view of the experimental setup to study G-APD optical readout from a THGEM-based two-phase Ar avalanche detector.

A G-APD (MRS APD "CPTA 149-35", [28]) was placed at a distance of 4 mm behind the second THGEM and 4 mm off-center of the frame. Accordingly, the G-APD viewing angle of the THGEM active-area, defined in Fig. 2 (left), amounted to about ±70º. Due to such a large viewing angle, the G-APD solid angle with respect to the avalanche scintillation (in fact with respect to the hole of the second THGEM), defined in Fig. 2 (right), considerably varied; its average value was determined by Monte-Carlo simulation (see section 5.2). The G-APD was isolated from the THGEM high voltage electrodes with a borosilicate glass plate transparent in NIR. The G-APD was optimized for the green-red range; it had a 4.41 mm$^2$ active area, 1764 (42×42) pixels, capacitance of 150 pF [28] and PDE~15% at 800 nm: see Fig. 3 [29]. The quenching resistor of each pixel was measured to be 25 MΩ and 5.7 GΩ at room temperature and at 88 K respectively.

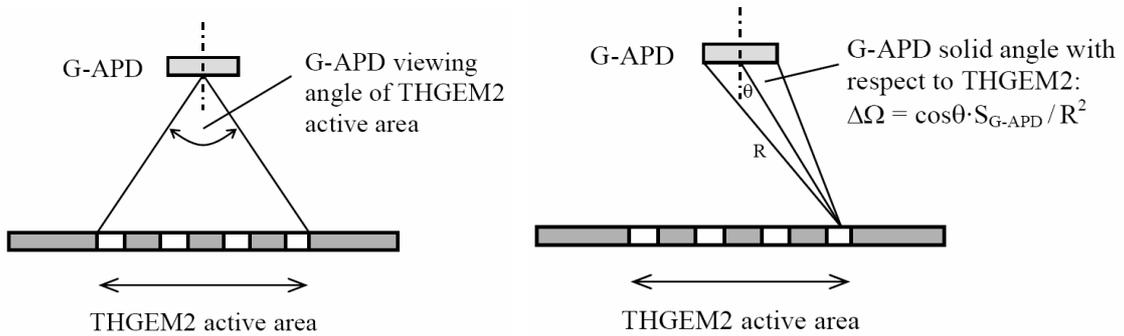

Fig. 2. Definition of the G-APD viewing angle of the THGEM2 active area (left) and the G-APD solid angle $\Delta\Omega$ with respect to the THGEM2 (right).

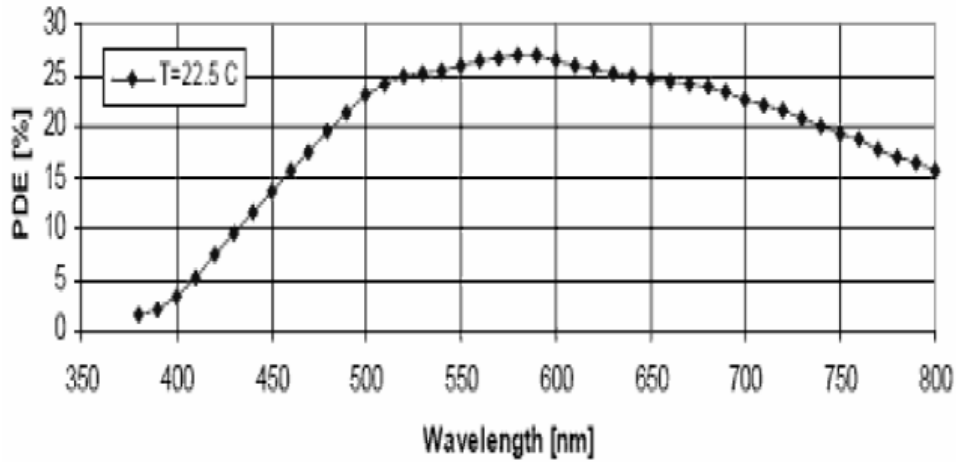

Fig. 3. Producer's specification [29] of the photon detection efficiency (PDE) of the G-APD (MRS APD "CPTA 149-35") used in this work.

The avalanche scintillation signal was read out from the G-APD via ~1 m long twisted-pair cable connected to a fast amplifier (CPTA, [28]) with 300 MHz bandwidth and amplification factor of 30; the amplifier was placed outside the cryogenic chamber. The signals were analyzed with a TDS5032B digital oscilloscope. The G-APD signal amplitude was expressed in photoelectrons (pe): the raw amplitude value was divided by single photoelectron (single pixel) amplitude, obtained in the course of studying G-APD single-pixel characteristics at cryogenic temperatures (see section 3).

The radiation-induced avalanche (charge) signal was read out from the last electrode of the second THGEM using a charge-sensitive amplifier with a shaping time of either 0.5 or 10 μs; the amplifier was placed outside the cryogenic chamber. In the following, the THGEM signal amplitude will often be expressed in initial electrons (e), i.e. in number of primary ionization electrons emitted into the gas phase from the liquid, prior to avalanche multiplication, using the amplifier sensitivity calibration (7.8 V/pC).

The signals in the detector were induced by either 60 keV X-rays from a $^{241}$Am source (with a frequency of few tens Hz) or 20-40 keV X-rays from a pulsed X-ray tube (with a frequency of 200 Hz); the detector was irradiated from outside, practically uniformly across the THGEM active area, through aluminium windows at the chamber bottom.

The THGEM charge gain was measured with pulsed X-rays with an amplifier shaping time of 10 μs, similar to that in our previous works [12],[13],[14],[15]: the gain is defined as the pulse-height of the avalanche (anode) signal of the THGEM divided by that of the calibration signal. The latter was recorded at the first electrode of the first THGEM, the cathode gap being operated in an ionization collection mode. The reproducibility of the gain value at a given voltage in the two-phase mode was about 10%.

## 3. G-APD characteristics in two-phase Ar

The G-APD gain, noise-rate and photon detection efficiency (PDE) characteristics at cryogenic temperatures are given elsewhere [24]; the results at 87 K are summarized as follows. Here we defined the G-APD over-voltage as the difference between the bias and the breakdown voltage, the latter being defined at the intersection of the gain-voltage characteristic with the voltage axis. Compared to room temperature, the breakdown voltage decreased at 87 K: from 34.5 to 30 V respectively. The typical linear growth of the pixel amplitude with voltage was found to end by its saturation at the bias voltage of 44 V, corresponding to an over-voltage of 14 V. The noise-rate considerably decreased with the temperature decrease: at 87 K it was as low as 700 and 1 Hz at bias voltages of 44 and 38 V respectively. The PDE reached a plateau at 35 V. In the following investigations the operation voltage of the G-APD was set at 44 V: at this value the signal amplitude before saturation reached a maximum, the PDE was on the plateau and the noise-rate was high enough to measure single-pixel characteristics.

In this work we further studied G-APD performance at cryogenic temperatures: we measured G-APD single-pixel characteristics and estimated a pixel dead-time by measuring the pixel quenching resistor.

Figs. 4 and 5 illustrate the G-APD single-pixel characteristics with the chosen 44V operation voltage, at 87 K. Fig. 4 shows typical single-pixel and triple-pixel noise signals; their major part has a width of 20 ns, reflecting the time-structure of the "Geiger" discharge in the pixel. In addition, the signal has a longer opposite-polarity tail, reflecting the characteristic response of the fast amplifier for the particular G-APD. For pure convenience, the original bipolar pulses were transformed to unipolar ones by integration, corresponding to pulse-shaping with a time constant of ~100 ns. The resulting pulse-area provided the G-APD signal amplitude. This permitted assessing the total amplitudes of long (~10 μs) G-APD signals, consisting of multiple short pulses separated in time, by signal integration over time. Such long signals will be considered in section 5 when studying avalanche scintillations in two-phase Ar.

Fig. 5 shows the G-APD noise spectrum; distributions of single-, double- and triple-pixel signals are well resolved, providing a well-defined single-pixel amplitude, used below to express the G-APD scintillation signal amplitude in number-of-photoelectrons. The spectrum of Fig. 5 also permits evaluating the cross-talk between pixels; the latter is defined as the ratio of the number of events with amplitude larger than 1 pixel to that with single-pixel amplitude. The cross talk value, averaged over several measurements at the bias voltage of 44 V, was found to be 41±10%. Therefore, in light-yield calculations, the measured G-APD amplitude should be corrected for this cross-talk (e.g. dividing by a factor of 1.41).

It should be remarked that we did not observe any indications of G-APD after-pulses in time scales of up to 10 ms.

The G-APD pixel dead-time after discharge is determined by the charging-up time, defined by the $R_Q C_P$ time-constant, where $R_Q$ and $C_P$ are the pixel quenching resistor and the pixel capacitance, respectively. In order to estimate $R_Q$, the G-APD current-voltage characteristics were measured in the forward direction at different temperatures: see Fig. 6. The slope of the I-V curve in its linear part corresponds to the total quenching resistor value R, being the sum of 1764 pixel quenching resistors connected in parallel; thus the pixel quenching resistor is $R_Q=1764 \cdot R$. Its dependence on temperature is illustrated in Fig. 7: the value $1/R_Q$ is shown as a function of $1/T$; in these variables, the Boltzmann-type temperature dependence of the conductivity $\sigma \sim \exp(-E_a/k_B T)$ (with $E_a$ being the activation energy of the conductivity) is better revealed.

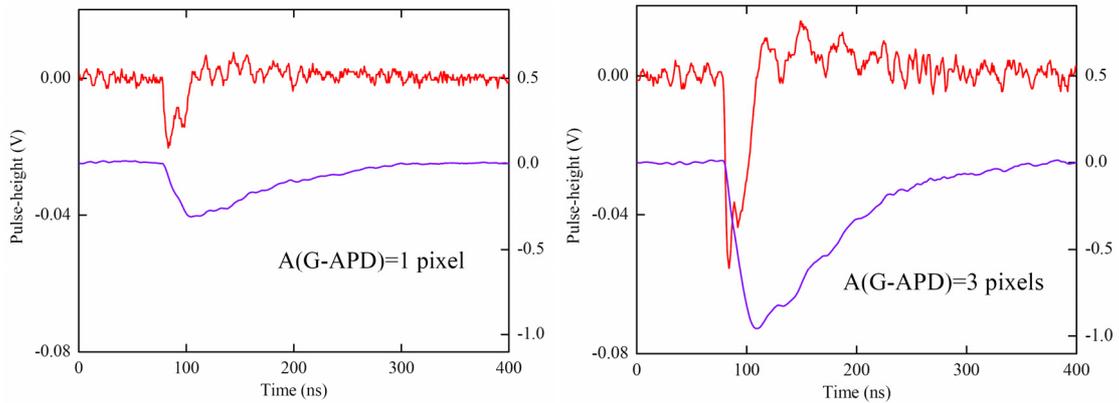

Fig. 4. Typical G-APD single-pixel (left) and triple-pixel (right) noise signals in two-phase Ar at 87 K, at the bias voltage of 44 V. Shown are the original bipolar pulses at the fast amplifier output (upper traces) and the unipolar pulses obtained after integration (lower traces).

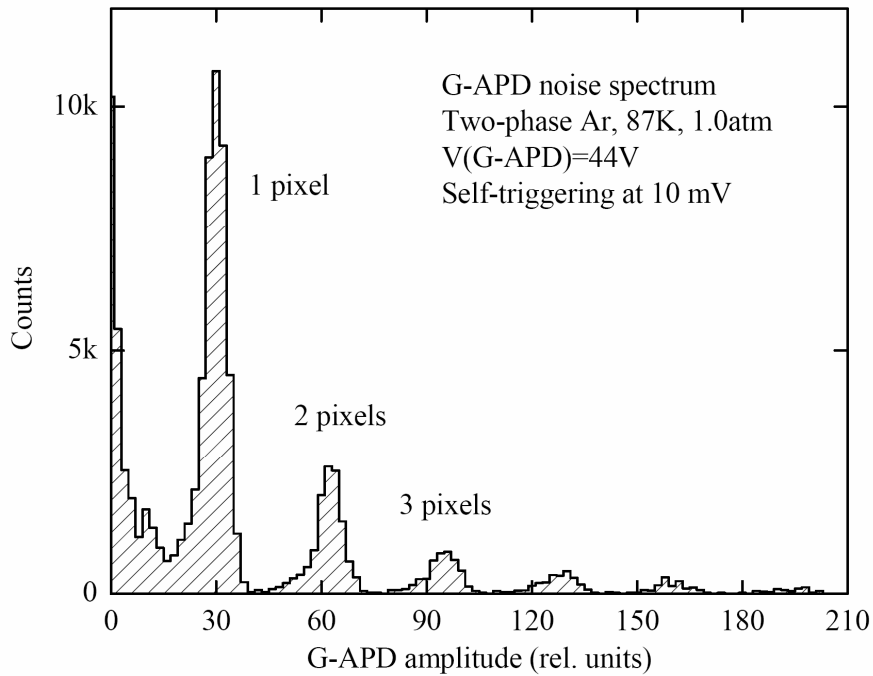

Fig. 5. G-APD amplitude noise spectrum in two-phase Ar at 87 K at the bias voltage of 44 V. The amplitude was obtained from the area under the G-APD unipolar pulses, integrated on a 400 ns scale.

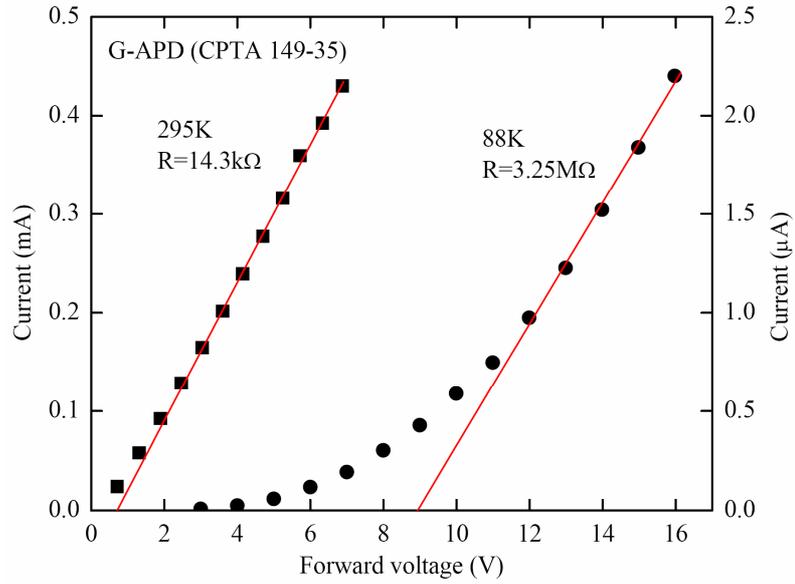

Fig. 6. G-APD current-voltage characteristic in the forward direction at room temperature (left scale) and at 88 K (right scale). The slope of the linear part of the I-V curve is defined by the G-APD total quenching resistor R.

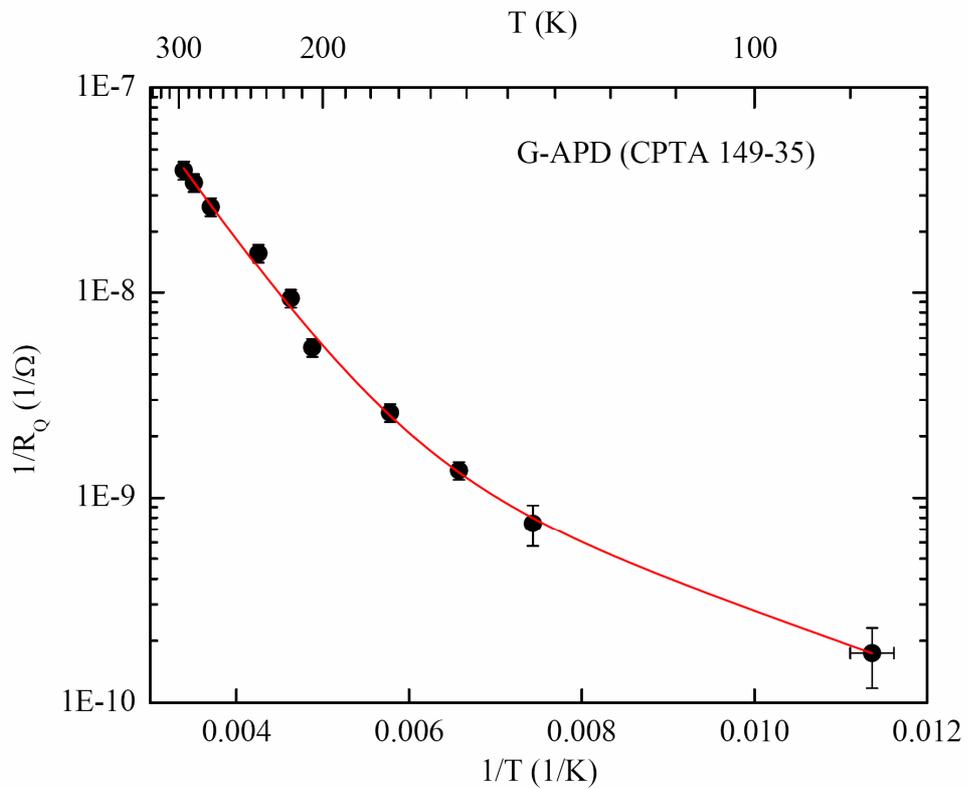

Fig. 7. G-APD single-pixel quenching resistor ($R_Q$) dependence on temperature: the reciprocal value $1/R_Q$ is shown as a function of the reciprocal temperature $1/T$.

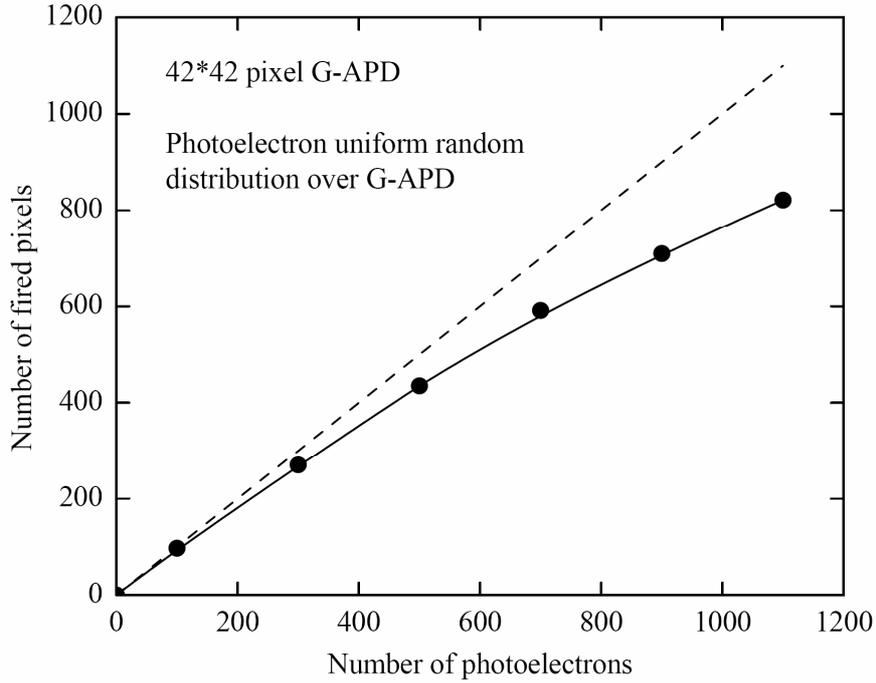

Fig. 8. Number of fired pixels in the G-APD with a matrix of 42×42 (1764) pixels as a function of the number of photoelectrons, obtained by Monte-Carlo simulation using uniform random distribution of photoelectrons over the G-APD area.

One can see that the quenching resistor dramatically increases with temperature decrease, by more than two orders of magnitude: the single-pixel quenching resistor varies from 25 M$\Omega$ at room temperature to 5.7 G$\Omega$ at 88 K. Taking into account the pixel capacitance $C_P$=150pF/1764=85 fF, the pixel dead-time constant is estimated to be of the order of 2 $\mu$s and 500 $\mu$s at room temperature and at 88 K respectively. The latter means that the rate capability of the entire G-APD in two-phase Ar in single-photoelectron counting mode is limited to a value of ~1764/500$\mu$s=3 MHz, which does not impose limitations in the applications foreseen. When having several photoelectrons per event, the rate capability will decrease inversely proportionally to the number of photoelectrons. In extreme cases the rate capability of the entire G-APD in two-phase Ar will be limited to ~1/500$\mu$s=2 kHz, which is certainly not a limiting factor in rare-event experiments.

Due to the rather large pixel dead-time at cryogenic conditions, one has to take into account the probability of multiple photoelectrons to hit the same pixel, leading to a nonlinear G-APD response. This is illustrated in Fig. 8, showing the number of activated pixels as a function of the number of photoelectrons induced per event, simulated for a G-APD with a matrix of 42×42 pixels. E.g., for 900 deposited photoelectrons only 710 will be detected by the G-APD, the others hitting already "fired" pixels. This correction will be used in section 5.2 to reconstruct the real photoelectron yield of the G-APD in the two-phase Ar avalanche detector operating at high gains.

## 4. THGEM avalanche charge characteristics in two-phase Ar

Figs. 9 and 10 illustrate the THGEM performance in two-phase Ar. Fig. 9 shows the gain-voltage characteristic of the double-THGEM (2THGEM) multiplier. Avalanche (charge) gains, reaching 1500, were obtained in two-phase Ar. Asymmetrical mode of operation of the 2THGEM was chosen, with the first THGEM biased to have a gain of 2 and the second THGEM voltage being varied. This resulted in the G-APD being sensitive to the avalanche scintillations essentially from the second THGEM, which facilitates interpretation of the results, in particular the calculation of the G-APD solid-angle with respect to the scintillation source (second THGEM holes). At the same time the effective collection of initial electrons into the first THGEM holes was provided.

Fig. 10 shows the avalanche-charge distributions from the 2THGEM at a gain of 400 induced by X-rays from $^{241}$Am. The avalanche charge was recorded using either the pulse-height or the pulse-area at a shaping time of 10 µs or 0.5 µs respectively. The 60 keV X-ray peak has a symmetrical shape indicating upon the high purity of liquid Ar, with electron life-time substantially larger than 4 µs [14]. In the following sections we selected electronically (threshold) ionization signals originating only from 60 keV X-rays converted in the liquid Ar phase. Taking into account the charge amplifier calibration and the gain, this 60 keV peak corresponds to 900±100 initial electrons.

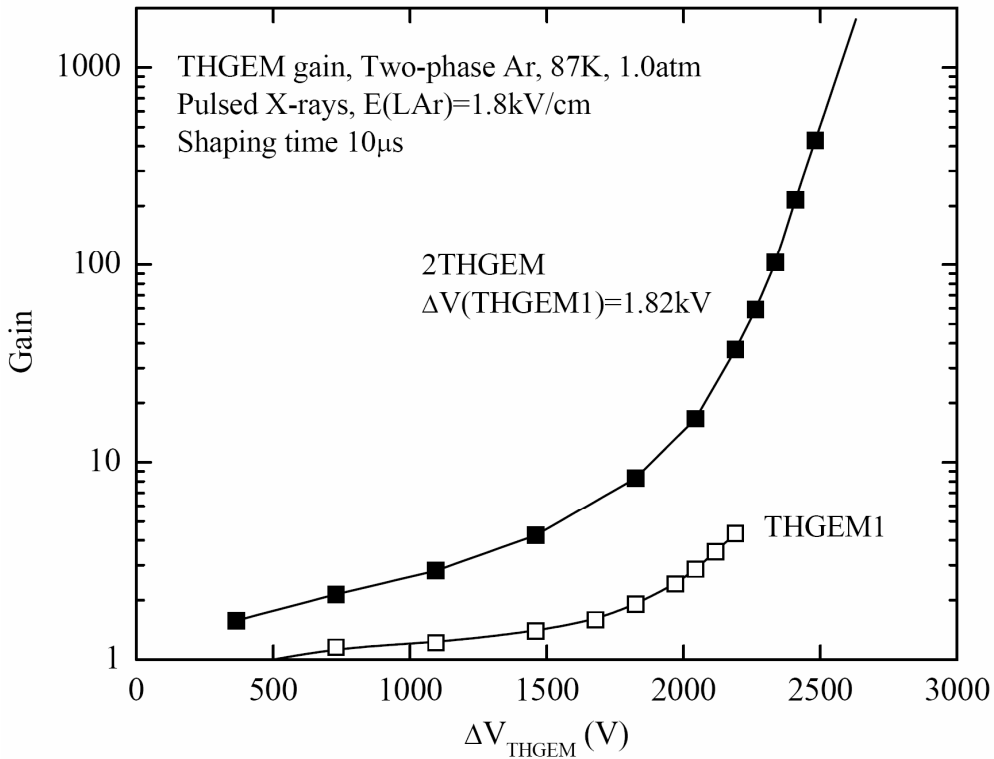

Fig. 9. Gain-voltage characteristic of the double-THGEM (2THGEM) multiplier at a fixed voltage of the first THGEM in two-phase Ar. The gain characteristic of the first THGEM (THGEM1) is also shown; note its gain of 2 at its given fixed voltage of 1.82 kV.

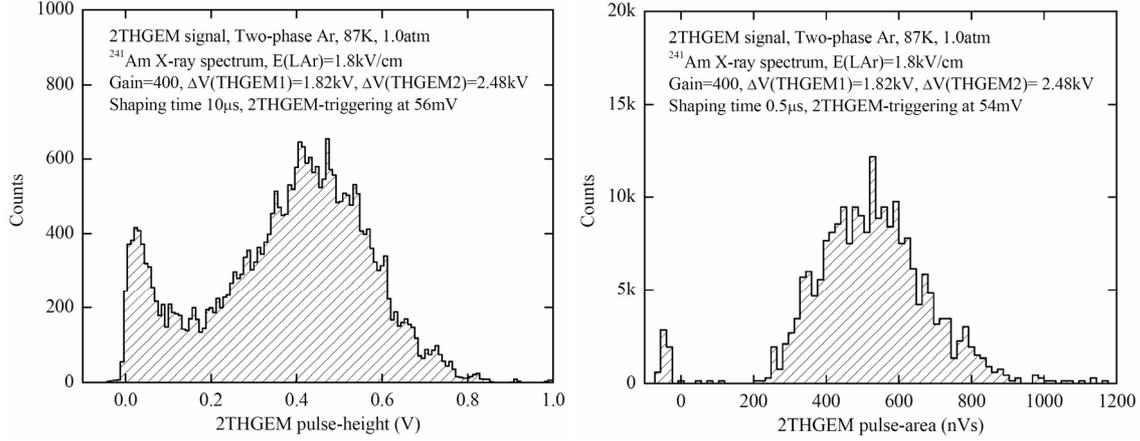

Fig. 10. Avalanche-charge distributions from the 2THGEM multiplier in two-phase Ar induced by X-rays from $^{241}$Am source, at charge amplifier shaping time of 10 μs using the pulse-height (left) and shaping time 0.5 μs using the pulse-area (right). 2THGEM gain of 400; electric field within the liquid of 1.8 kV/cm; 2THGEM self-triggering at a threshold of about 55 mV.

## 5. Detection of avalanche scintillations in two-phase Ar

### 5.1 General properties of avalanche-scintillation signals

We detected avalanche scintillations in the THGEM with a bare G-APD (without WLS, i.e. insensitive to UV), the scintillation being most probably in the NIR, as discussed above. The scintillation was observed both in the two-phase mode at cryogenic temperature and in gaseous Ar at room temperature.

The correlation between scintillation (G-APD) and charge (THGEM) signals is illustrated in Fig. 11 in gas Ar at room temperature and in two-phase Ar at 87 K; the correlation is indicated by arrows. While at both temperatures the slow charge signals overlap, the resulting correlated G-APD signals are fast and well resolved. Note that in Ar gas at room temperature, the avalanche signal is somewhat faster compared to that in the vapor phase above liquid Ar (the latter containing a slow electron-emission component [30]), which permits resolving closer (in time) ionization clusters (Fig. 11).

Fig. 12 depicts the correlation of the scintillation-versus-charge signals in two-phase Ar, induced by 60 keV X-rays absorbed in liquid Ar; the 2THGEM gain was set to 400. The relatively large amplitude fluctuations observed in the G-APD are due to that of the solid angle, as discussed in the next section.

It should be stressed that the rather fast G-APD response provided an effective means to study the avalanche signal shape and accordingly the electron emission and avalanche mechanisms in two-phase Ar. Typical avalanche-scintillation and avalanche-charge signals of the 2THGEM (at a gain of 400) in two-phase mode, induced by 60 keV X-rays absorbed in liquid Ar, are presented in Fig. 13; the G-APD signal amplitudes were of 430 and 1000 photoelectrons. The time structure of the scintillation pulses reflects the electron emission processes at the liquid-gas interface at the present electric fields [30]: the spike at the beginning of the pulse is induced by the fast electron emission component and the tail by the slow component, sometimes modulated by ion feedback-induced secondary avalanches.

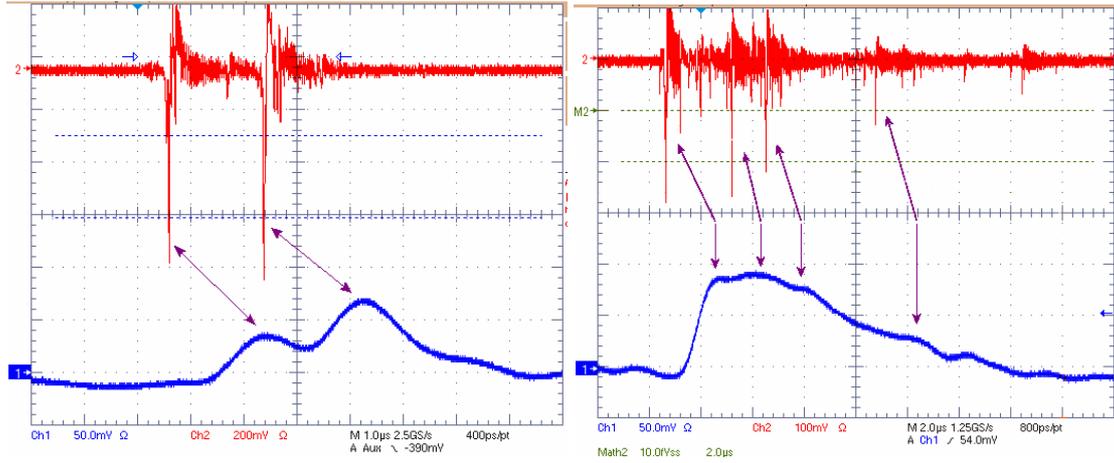

Fig. 11. Correlated avalanche-charge (2THGEM anode, lower traces) and avalanche-scintillation (G-APD, upper traces) signals; the correlation is indicated by arrows. Left: 1 atm Ar gas at 295 K; pulsed X-ray source; 2THGEM gain 18. Right: two-phase Ar at 87 K; 60 keV X-rays from $^{241}$Am source converted in liquid Ar; 2THGEM gain 400. G-APD bias voltage: 42 V (left) and 44 V (right); charge amplifier shaping time: 0.5 μs.

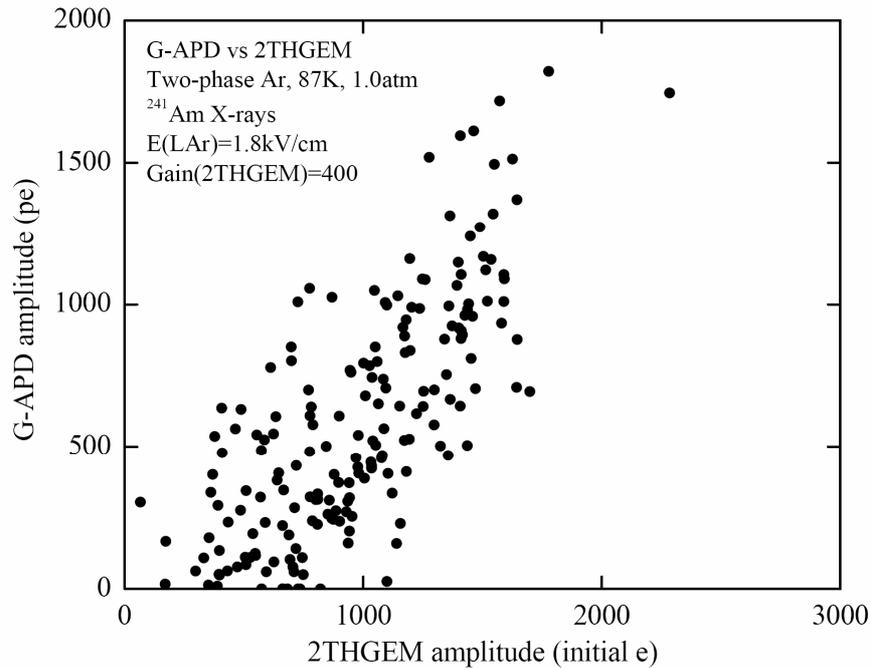

Fig. 12. Correlation between avalanche-charge and avalanche-scintillation signals. Shown is the G-APD scintillation-signal amplitude (expressed in photoelectrons, not corrected for nonlinearity and cross-talk) versus 2THGEM charge-signal amplitude (expressed in initial electrons). The signals of 60 keV X-rays from $^{241}$Am source converted in liquid Ar were induced in a 2THGEM operated in two-phase Ar. 2THGEM gain: 400; electric field within the liquid: 1.8 kV/cm; G-APD bias voltage: 44 V; charge-amplifier shaping time: 0.5 μs.

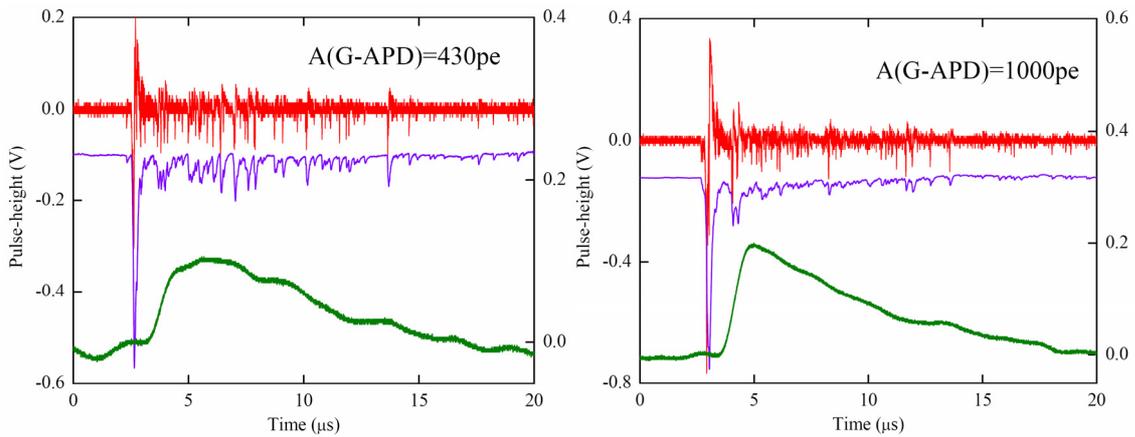

Fig. 13. Typical G-APD avalanche-scintillation signals (upper traces for bipolar pulse and middle traces for slightly filtered unipolar pulse; left scale) and 2THGEM avalanche-charge signals (lower traces; right scale). The signals of 60 keV X-rays from $^{241}$Am source converted in liquid Ar were induced in a 2THGEM operated in two-phase Ar at 87 K. 2THGEM gain: 400; G-APD bias voltage: 44 V; electric field within liquid Ar: 1.8 kV/cm; charge-amplifier shaping time: 0.5 μs. The integrated G-APD amplitude is 430 photoelectrons (left) and 1000 photoelectrons (right); numbers were not corrected for nonlinearity and cross-talk.

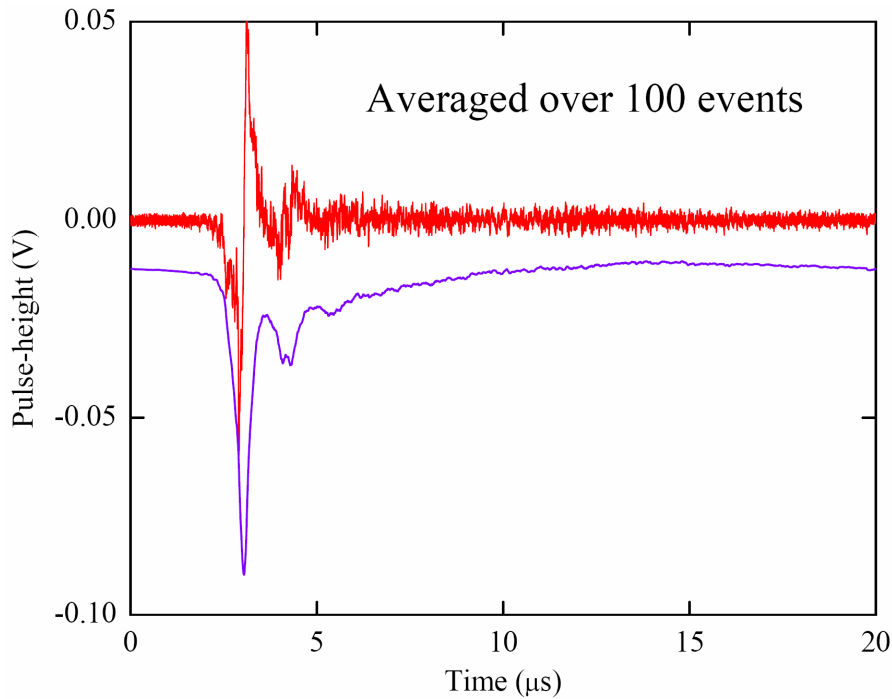

Fig. 14. Typical G-APD avalanche-scintillation signals averaged over 100 events: bipolar (upper trace) and slightly filtered unipolar (lower trace). The signals of 60 keV X-rays from $^{241}$Am source converted in liquid Ar were induced in a 2THGEM operated in two-phase Ar at 87 K. 2THGEM gain: 400; G-APD bias voltage: 44 V; electric field within liquid Ar: 1.8 kV/cm.

The G-APD signal's time structure is better revealed when averaging signals, as shown in Fig. 14: the presence of the fast (the pulse spike) and slow (the pulse tail) components of the electron emission through the liquid-gas interface [30] is evident. This pulse shape strongly resembles that of the avalanche-charge signal observed elsewhere (Fig. 3 in [30]) when studying electron emission properties in two-phase Ar at similar emission field, using faster GEM multipliers.

The tail of the averaged pulse is modulated by two cycles of ion feedback between the electrodes of the second THGEM (Fig. 14). From this one can estimate an ion feedback cycle of ~1.2 μs, corresponding to the ion drift time though the THGEM holes and therefore resulting in the relatively slow signals (induced mostly by ion drifting through the hole). Accordingly, the observed fast and slow components cannot be resolved with the charge-signal readout (see Fig. 13).

From Fig. 14 one could quantitatively estimate the ion-feedback contribution to the avalanche: at a 2THGEM gain of 400 it amounted at ~15% of that of the primary signal. At lower gains, e.g. 60 and 4, ion feedback was not observed.

The time constant of the slow component can be measured from the time distributions shown in Fig. 15; they were obtained by recording the negative part of the waveforms of the G-APD bipolar signals, with respect to the THGEM trigger. Indeed, due to the rather fast G-APD single-pixel pulse (see Fig. 4), the tail of the G-APD avalanche scintillation signal is composed of many non-overlapping single-pixel pulses of a standard shape (see Figs. 13). Their time distribution will thus be correctly reflected by the time histogram of Fig. 15, except for the initial part of the signal (first few μs) where the single-pixel pulses overlap.

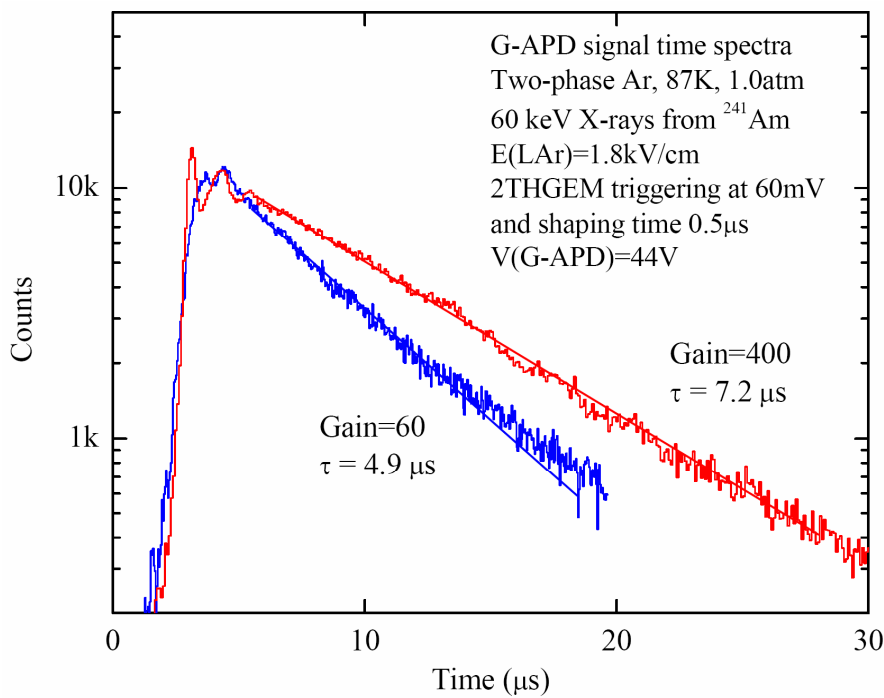

Fig. 15. Time distributions obtained by recording the negative part of the waveforms of the G-APD avalanche-scintillation bipolar signals, with respect to the 2THGEM trigger. The signals of 60 keV X-rays from $^{241}$Am source converted in liquid Ar were induced in a 2THGEM operated in two-phase Ar at 87 K. 2THGEM gains: 60 and 400; G-APD bias voltage: 44 V; electric field within liquid Ar: 1.8 kV/cm.

We did not observe any difference in time distribution between 2THGEM gains of 4 and 60; at these moderate gains the signal tail, defined by slow electron emission component in two-phase Ar, had a decay time of 4.9 μs (at an electric field in liquid Ar of 1.8 kV/cm); see Fig. 15. This value is in accordance with that obtained earlier with the avalanche-charge signal of the somewhat faster GEM multiplier [30]: 5.3±1 μs at 1.7 kV/cm. Note that at gain 400 the avalanche signal was slower, with decay time of 7.2 μs (Fig. 15), in accordance with our previous observations [15].

The good quantitative agreement between the slow-component time constant of scintillation signal measured here and that of the charge signal measured elsewhere [30] may suggest that the time constant of the avalanche scintillation within the G-APD sensitivity range (presumably NIR) is below the sub-microsecond scale; i.e. NIR scintillations are much faster compared to those of VUV (the latter having time constants of about 3 μs [31]).

## 5.2 THGEM/G-APD multiplier yield and avalanche-scintillation light yield

Fig. 16 characterizes the combined THGEM/G-APD multiplier yield: the avalanche-scintillation signal amplitude distribution of 60 keV X-rays, in two-phase Ar, is shown at a 2THGEM gain of 400; the G-APD signals were integrated over 20 μs. The raw average value of 710 photoelectrons corrected for nonlinearity of the G-APD (section 3, Fig. 8), corresponds to 900 photoelectrons. This includes both the direct scintillation signal and the secondary cross-talk one, discussed above. A further correction by the cross-talk factor of 1.4 (section 3) results in an absolute yield of scintillation-induced 640±40 photoelectrons. This value originates from about 900±100 initial electrons produced by a 60 keV X-ray converted in liquid Ar (section 4). Consequently, at this particular THGEM gain, electric field in the liquid and solid angle (see below), the combined THGEM/G-APD yield, i.e. the average number of photoelectrons produced directly by avalanche-scintillation photons per *initial electron* or per *deposited energy*, amounts at:

$$Y_{THGEM/G-APD}(400) = 0.71 \pm 0.09 \, pe/initial \ e = 11 \pm 1 \, pe/keV$$

Note that these values were obtained at a rather large G-APD viewing angle, roughly at ±70°, the viewed area (the THGEM active area) being practically uniformly irradiated with the ionization. This resulted in a rather small G-APD solid angle with respect to the second THGEM; simulations yielded an average solid angle $\Delta\Omega/4\pi = 2.7 \times 10^{-3}$ (Fig. 17).

Nevertheless, the THGEM/G-APD yield turned out to be relatively large even at this non-optimized geometry. In particular, for an effective operation in single-electron counting mode (with a statistical request imposed by detection efficiency of > 3 pe per initial electron), the solid angle, THGEM gain and detected light-yield should be jointly increased by a reachable factor of 5. For example, calculations indicated that the average solid angle could be dramatically increased, by a factor of 5, simply reducing the viewing angle down to more practical value, namely down to ±45°.

The avalanche-scintillation light yield can be better estimated from Fig. 18, showing the G-APD-to-THGEM amplitudes ratio distribution, with both amplitudes being expressed in respective numbers of photoelectrons and avalanche-electrons. The distribution is rather broad: it reflects that of the G-APD solid angle with respect to the second THGEM (to compare with Fig. 17). Taking the raw average value (0.002 pe/e), correcting for nonlinearity and cross-talk and taking into account the G-APD average solid angle, we obtained the following avalanche-scintillation photoelectron-yield extrapolated to 4π acceptance, as measured with the G-APD (at a gain of 400):

$$Y_{PE}(400) = 0.66 \pm 0.2 \, pe / avalanche \ e$$

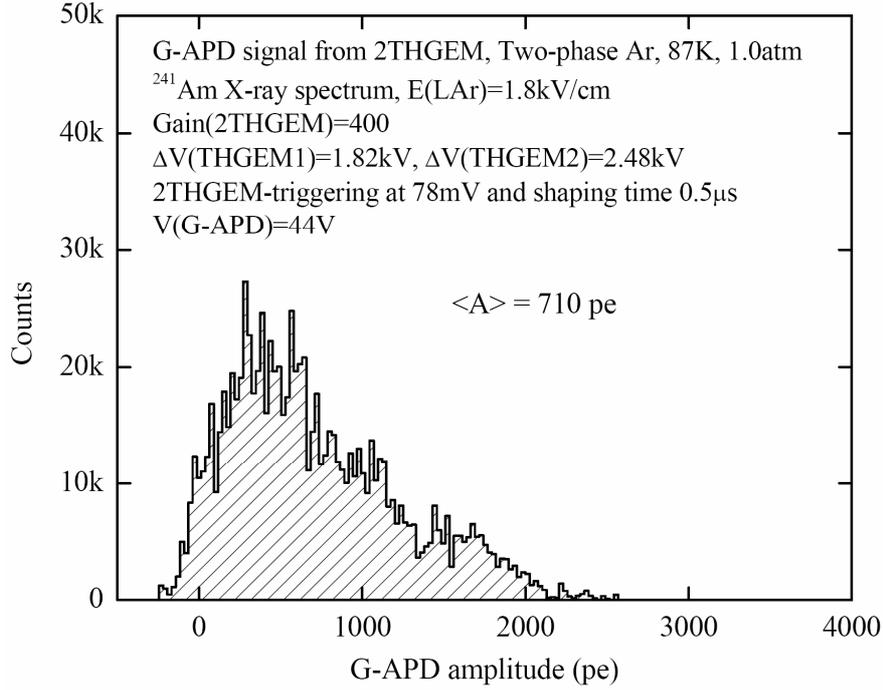

Fig. 16. Avalanche-scintillation amplitude spectrum from the 2THGEM read out by the G-APD. The amplitude is expressed in G-APD photoelectrons (pe) integrated over 20 μs, not corrected for nonlinearity and cross-talk. The signals of 60 keV X-rays from $^{241}$Am source converted in liquid Ar were induced in a 2THGEM operated in two-phase Ar at 87 K. 2THGEM gain: 400; G-APD bias voltage: 44 V; electric field within liquid Ar: 1.8 kV/cm. The raw average amplitude is 710 pe.

Furthermore, accounting for the G-APD PDE of 15% at 800 nm, the avalanche-scintillation photon yield over 4π can be estimated:

$$Y_{PH}(400) = 4.4 \pm 1 \, ph / avalanche \ e$$

This is a rather high light yield; it should be compared to that of ~1 photon per avalanche electron presented in [27] for avalanche scintillations in Ar in the NIR region.

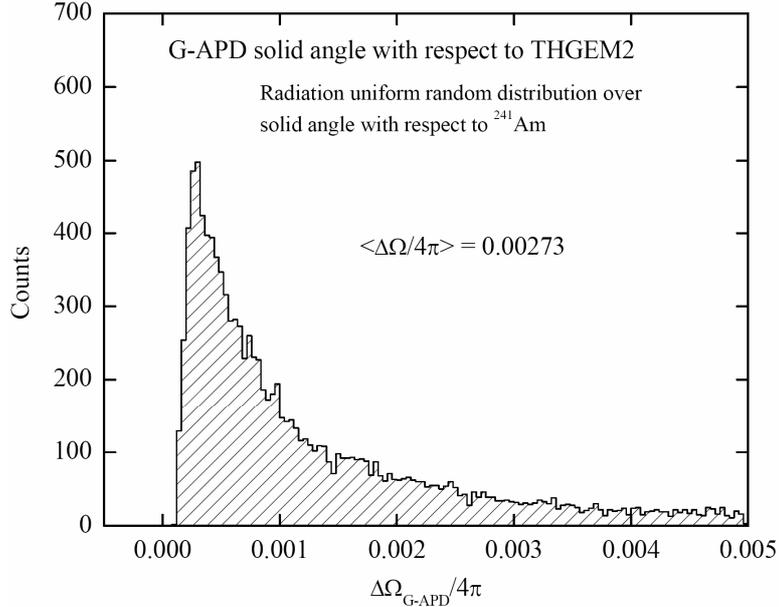

Fig. 17. Distribution of the G-APD solid angle with respect to the active area of the second THGEM, obtained by Monte-Carlo simulation using uniform random distribution of the radiation over the solid angle with respect to $^{241}$Am source, resulting in practically uniform ionization distribution over THGEM area. The light refraction in the glass plate in front of the G-APD was taken into account, the light reflection being neglected.

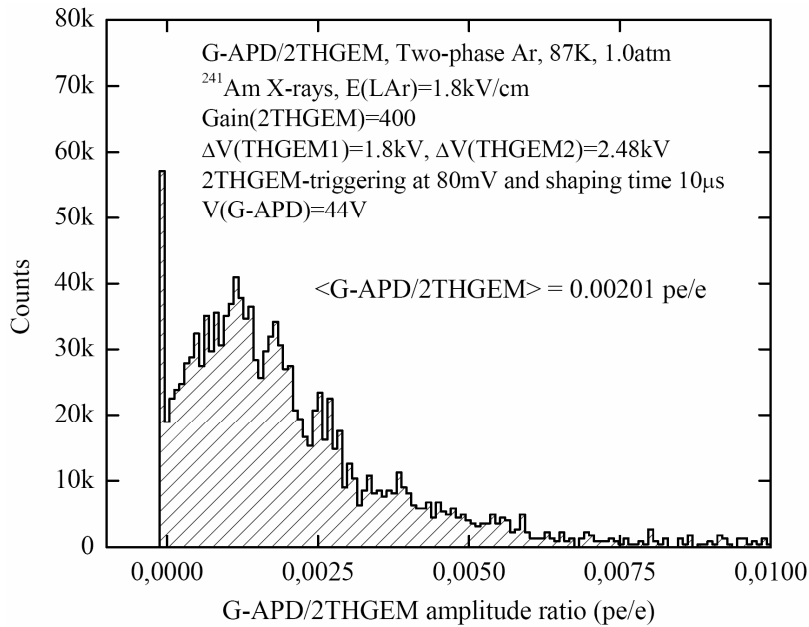

Fig. 18. Avalanche scintillation-to-charge signals ratio spectrum. The ratio is expressed as that of the number of G-APD photoelectrons, not corrected for nonlinearity and cross-talk, to the number of avalanche electrons as recorded by the 2THGEM. The signals of 60 keV X-rays from $^{241}$Am source converted in liquid Ar were induced in a 2THGEM operated in two-phase Ar at 87 K. 2THGEM gain: 400; G-APD bias voltage: 44 V; electric field within liquid Ar: 1.8 kV/cm. The raw average ratio is 0.0020 pe/e.

| Double-THGEM gain | 400 | 60 |
|---|---|---|
| THGEM/G-APD yield, photoelectron | 640 | 130 |
| THGEM/G-APD yield, photoelectron per initial e | 0.71±0.09 | 0.14±0.02 |
| THGEM/G-APD yield, photoelectron per keV | 11±1 | 2.1±0.3 |
| THGEM/G-APD reduced yield over 4π, photoelectron per keV per unit gain | 9.8±3 | 13±4 |
| Light yield over 4π, photoelectron per avalanche electron | 0.66±0.2 | 0.88±0.3 |
| Light yield over 4π, NIR photon per avalanche electron | 4.4±1 | 5.6±2 |

Table 1. A summary of THGEM/G-APD yields (first five rows) and avalanche-scintillation light yields (last two rows) for 2THGEM gains of 400 and 60, measured in two-phase Ar. 60 keV X-rays from $^{241}$Am were converted in liquid Ar. Electric field within the liquid: 1.8 kV/cm; G-APD viewing angle: ±70º. The data were corrected for G-APD nonlinearity and cross-talk.

The THGEM/G-APD multiplier yields and the avalanche-scintillation light yields at a 2THGEM gain of 400 are summarized in Table. 1. Very similar data measured at a lower gain of 60 [24] are also presented in the table.

To compare our results to those obtained at different gains and by other authors, we also included in Table 1 the reduced THGEM/G-APD yield over 4π; it is expressed in photoelectrons per keV of deposited energy per unit gain of the 2THGEM. The values presented, 10-13 pe/keV at gains of 60-400, should be compared to that of 7 pe/keV over 4π per unit gain, deduced from recent data of [16], obtained at THGEM gain 100 with a G-APD coated with a WLS. It is interesting to note that these very similar results were obtained in different systems over different wavelength ranges: NIR and VUV.

## 6. Conclusions

A novel concept of optical signal recording in two-phase avalanche detectors, with G-APD measuring avalanche-scintillation photons in a THGEM, has been studied in view of its potential applications in rare-event experiments and other fields.

The direct detection of avalanche scintillations from THGEM holes in two-phase Ar has been demonstrated with a bare G-APD uncoated with wavelength shifter, i.e. insensitive in the UV. The high sensitivity and fast response of the G-APD provided an effective means for observing electron emission and avalanche phenomena at two-phase Ar conditions. In particular, the fast and slow components of electron emission through the liquid-gas interface as well as ion feedback-induced avalanches in THGEM holes were effectively visualized with the THGEM/G-APD multiplier.

The combined THGEM/G-APD light-yields at an avalanche gain of 400 and under G-APD viewing angle of ±70º were measured to be 640 photoelectrons (pe) per 60 keV X-ray converted in liquid Ar; this corresponds to ~0.7 pe per initial deposited (prior to avalanche-multiplication) electron.

The avalanche scintillations observed occurred presumably in the near infrared (NIR) where G-APDs have high sensitivity; these scintillations are relatively fast having sub-microsecond time constant. In THGEM multiplication-mode the avalanche-scintillation light-yield measured by the G-APD was 0.7-0.9 pe per avalanche-electron extrapolated to 4π acceptance, at a gain of 60-400; this would correspond to NIR photon yield of about 5 photons per avalanche electron over $4\pi$. Such a high avalanche-scintillation light yield, similar to that observed in the VUV with WLS-coated G-APDs, clearly indicates upon the applicability of the optical readout of two-phase Ar avalanche detectors with THGEM multipliers viewed by G-APDs.

It should be remarked that similar optical readout, with present bare G-APDs sensitive in NIR, cannot be yet applied in two-phase detectors with noble gases other than Ar. In particular as concerns Xe, its NIR emission is around 1300 nm [32], where silicon is transparent and silicon-based photon detectors thus cannot be used. Some photodiodes made of other materials, e.g. InGaAs, are sensitive in this region but cannot operate presently in single-photoelectron counting mode. To our best knowledge, there is no precise information regarding NIR emission of other noble gases. On the other hand, UV-sensitive WLS-coated G-APDs could be applicable to detectors operating with Ar, Xe, Kr and Ne.

A practical optical readout of two-phase Ar avalanche detectors would comprise G-APDs matrices placed a few millimeters behind THGEM multipliers, with a pitch of ~1 cm, viewing clusters of multiplier holes under an angle of ±45º; these would cover the detector's active area with spatial resolutions sufficient for rare-event experiments. For example, for a 100 kg liquid Ar TPC of a volume of 40×40×40 cm$^3$ the total number of G-APDs would be reasonable – of 1600. Such a detector would be robust, stable, simple and relatively cheap. Further studies are in progress, including that related to the natural radioactivity of the detector elements.

## 7. Acknowledgements

We are grateful to Y. Tikhonov for the support, R. Snopkov and A. Chegodaev for the development of the experimental setup, M. Danilov and E. Kravchenko for providing with G-APDs, A. Akindinov, J. Haba and Y. Kudenko for the discussion of the results. This work was supported in part by RFFI grant 09-02-12217-ofi_m and by Federal special program "Scientific and scientific-pedagogical personnel of innovative Russia" in 2009-2013. A. Breskin is the W.P. Reuther Professor of Research in the peaceful use of Atomic Energy.**References**

[1] B.A. Dolgoshein et al., *New method of registration of ionizing-particle tracks in condensed matter*, JETP Lett. 11 (1970) 513.

[2] A. Buzulutskov et al., *First results from cryogenic avalanche detectors based on gas electron multipliers*, IEEE Trans. Nucl. Sci. NS-50 (2003) 2491.

[3] A. Buzulutskov, *Radiation detectors based on gas electron multipliers (Review)*, Instr. Exp. Tech. 50 (2007) 287, and references therein.

[4] F. Sauli, *GEM: A new concept for electron amplification in gas detectors*, Nucl. Instrum. Meth. A 386 (1997) 531.

[5] A. Breskin et. al., *A concise review on THGEM detectors*, Nucl. Instr. Meth. A 598 (2009) 107.

[6] A. Buzulutskov et al., *The GEM photomultiplier operated with noble gas mixtures*, Nucl. Instrum. Meth. A 443 (2000) 164.– 18 –